\documentclass[cits]{PoS}
\usepackage{amsmath}
\usepackage{amssymb}
\usepackage{bm}
\usepackage{slashed}
\usepackage{mathtools}
\usepackage{cases}

\title{\vspace*{-6.5cm}
{\hfill \texttt{\footnotesize CERN-PH-TH-2014-185}}
\vfill
Deformations of infrared-conformal theories in two dimensions}

\ShortTitle{Deformations of infrared-conformal theories in two dimensions}

\author{\speaker{Oscar Akerlund}$^a$ and Philippe~de~Forcrand$^{ab}$\\
        \llap{$^a$} Institut f\"ur Theoretische Physik, ETH Zurich, CH-8093 Z\"urich, Switzerland\\
        \llap{$^b$} CERN, Physics Department, TH Unit, CH-1211 Geneva 23, Switzerland\\
        E-mail: \email{oscara@itp.phys.ethz.ch}, \email{forcrand@itp.phys.ethz.ch}}

\abstract{We study two exactly solvable two-dimensional conformal models, the critical Ising 
model and the Sommerfield model, on the lattice. We show that finite-size effects are important and
depend on the aspect ratio of the lattice. In particular, we demonstrate how to obtain the correct massless
behavior from an infinite tower of finite-size-induced masses and show that it is necessary to first take
the cylindrical geometry limit in order to get correct results. 
In the Sommerfield model we also introduce a mass deformation to measure the mass anomalous dimension,
$\gamma_m$. We find that the explicit scale breaking of the lattice setup induces
corrections which must be taken into account in order to reproduce $\gamma_m$
at the infrared fixed point. These results can be used to improve the
methodology in the search for the conformal window in QCD-like theories with many flavors.}

\FullConference{The 32nd International Symposium on Lattice Field Theory,\\
		23-28 June, 2014\\
		Columbia University New York, NY}

\newcommand{\R}{\ensuremath{\mathbb{R}}}

\newcommand{\rd}{\ensuremath{\mathrm{d}}}
\newcommand{\idb}[1][x]{\ensuremath{\!\!\rd#1\,}}

\providecommand{\abs}[1]{\left\lvert#1\right\rvert}

\newcommand{\expv}[1]{\left\langle#1\right\rangle}
\newcommand{\bra}[1]{\langle#1\rvert}
\newcommand{\ket}[1]{\lvert#1\rangle}

\newcommand{\bracket}[3]{\bra{#1}#2\ket{#3}}

\newcommand{\be}{\begin{equation}}
\newcommand{\ee}{\end{equation}}
\newcommand{\benn}{\nonumber\begin{equation}}
\newcommand{\eenn}{\nonumber\end{equation}}

\newcommand{\floor}[1]{\ensuremath{\left\lfloor#1\right\rfloor}}

\def\bea{\begin{eqnarray}} \def\eea{\end{eqnarray}}
\def\beann{\begin{eqnarray*}} \def\eeann{\end{eqnarray*}}
\def\lsim{\raise0.3ex\hbox{$<$\kern-0.75em\raise-1.1ex\hbox{$\sim$}}}
\def\gsim{\raise0.3ex\hbox{$>$\kern-0.75em\raise-1.1ex\hbox{$\sim$}}}

\newcommand*\xbar[1]{%
  \hbox{%
    \vbox{%
      \hrule height 0.5pt % The actual bar
      \kern0.3ex%         % Distance between bar and symbol
      \hbox{%
        \kern-0.1em%      % Shortening on the left side
        \ensuremath{#1}%
        \kern-0.1em%      % Shortening on the right side
      }%
    }%
  }%
}

\begin{document}

\section{Introduction}\label{sec:introduction}\noindent
(Near-)conformal theories, for example Walking Technicolor, can be viable candidates for physics
beyond the standard model (BSM). For these models to be phenomenologically
relevant, large anomalous dimensions are required at the infrared fixed point which often cannot
be realized at weak coupling. Strongly coupled gauge theories are best studied on the lattice
but the lattice is a priori not well suited to study (near-)conformal theories because of explicit
breaking of rotational and scale invariance through the UV and IR cutoffs.
It is therefore of great importance to properly understand how a conformal theory is deformed by
the lattice in order to correctly interpret potential remnant signatures of its conformality.

\section{Deformed correlators}\label{sec:defcorr}\noindent
To understand how a nonzero mass deforms a conformal model it is useful to look at the (Euclidean) two-point
correlator of a conformal field with anomalous scaling dimension $\Delta$, i.e. with a coordinate
space two-point function $G(x) \propto (x^2)^{-\tfrac{d-2}{2}-\Delta}$, where $d$ is the space-time
dimension. Power counting tells us that the propagator in Fourier space is
$\widetilde{G}(p) \propto (p^2)^{-1+\Delta}$ from which the free, massless case $\Delta=0$ is
evident. One can introduce a mass deformation in the usual sense by shifting the
pole of the propagator from zero to $im$ for which $\widetilde{G}(p) = (p^2+m^2)^{-1+\Delta}$.
The zero spatial momentum euclidean time propagator is then given by a one-dimensional Fourier transform:
\begin{equation}
  G(t,\vec{p}=0) \propto \int\idb[\omega]\widetilde{G}\left(p = (\omega,\vec{0})\right)e^{-i\omega t} = \int\idb[\omega](\omega^2+m^2)^{-1+\Delta}e^{-i\omega t}\propto \left(\frac{t}{m}\right)^{\tfrac{1}{2}-\Delta}K_{\tfrac{1}{2}-\Delta}(mt)
\end{equation}
In the infrared ($mt\gg1$) an expansion of the Bessel function reveals that the usual
exponential decay due to the mass is modified by a power law in $t$:
\begin{equation}
  \label{eq:pole_shift}
  G(t,\vec{p}=0) \propto \frac{e^{-mt}}{t^{\Delta}}.
\end{equation}
This relation has been used in \cite{Iwasaki:2014} as a method to extract anomalous scaling dimensions.
However, it relies on a very strong assumption. Namely, the way to introduce a mass deformation is
not unique. It has been pointed out \cite{Stephanov:2007} that a
massless correlator with anomalous dimension $\Delta$ can be viewed as a field with a continuous
mass distribution given by a power law with parameter $(\Delta-1)$, since
\begin{equation}
  \label{eq:unparticle}
  \left(p^2\right)^{-1+\Delta} = \frac{\sin\Delta\pi}{\pi}\int_0^\infty\idb[M^2]\frac{\left(M^2\right)^{\Delta-1}}{p^2+M^2}.
\end{equation}
Then, a finite mass deformation can be introduced by making the mass distribution $(M^2)^{\Delta-1}$ discrete with
a finite mass spacing. The correlator will then change from a power law to an infinite sum of
exponentials, which is conceptually different from a single exponential with a power correction.
This second description offers a smooth transition from a scale invariant correlator to the familiar
exponential correlator of a gapped system which is not present in the first. The second view will also
be shown to be the correct one in the analysis of the critical Ising model below.

\section{The $2d$ critical Ising model}\label{sec:2dising}
\begin{figure}
  \centering
    \mbox{\includegraphics[width = 0.42\textwidth]{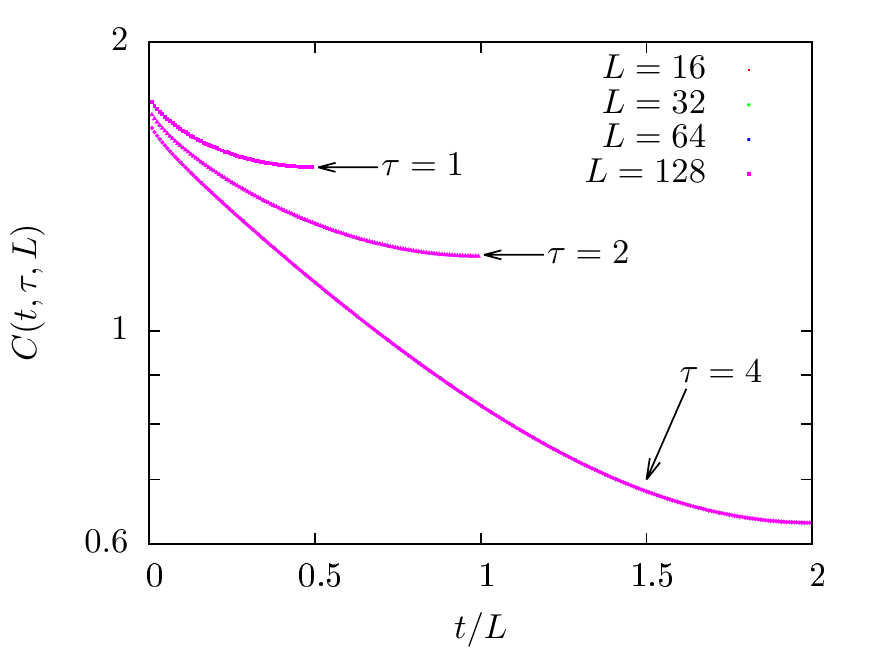}%
      \includegraphics[width = 0.42\textwidth]{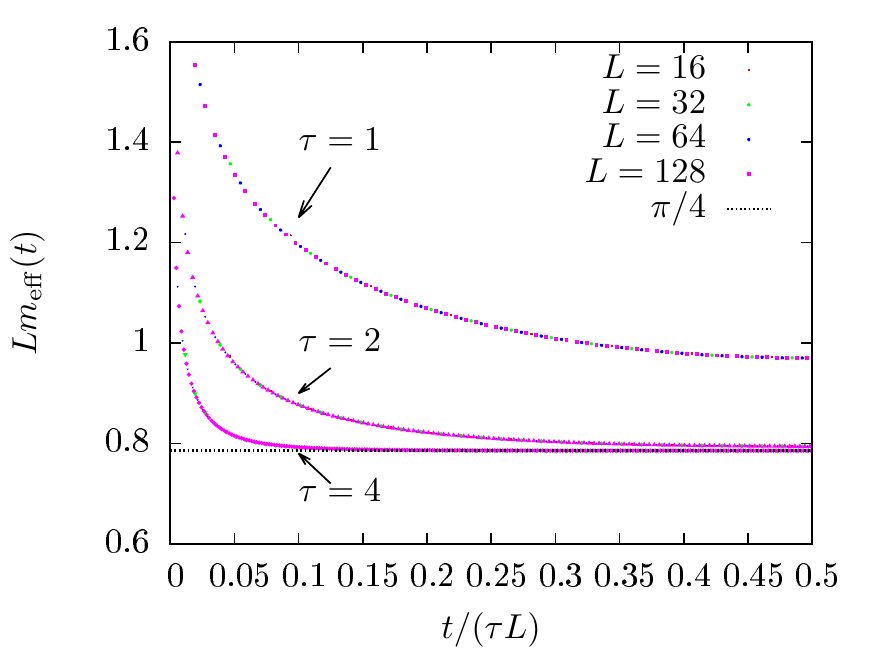}}
    \caption{Zero-momentum correlators (\emph{left}) and effective masses (\emph{right}) of the critical $2d$ Ising model
      for various lattice sizes $L\times(\tau L)$.
      Note the false plateau of the effective mass far away from the lowest mass for $\tau=1$.}
    \label{fig:ising_meff}
  %\end{center}
\end{figure}

\noindent
The $2d$ critical Ising model is \emph{the} textbook example of a conformal field theory.
Everything about this model is known and the analytic expression for the two-point function on a
torus is particularly useful for our purposes. However, first recall the form of the two-point
function in $\R^2$:
\begin{equation}
  \label{eq:ising_r2}
  \expv{\sigma(0,0)\sigma(x,t)} \propto \left(x^2+t^2\right)^{-\tfrac{1}{8}}
\end{equation}
where the power law decay is governed by the scaling dimension $\Delta=\tfrac{1}{8}$ of the periodic spin
field. On a torus  of size $L_s\times L_t\equiv L\times(\tau L)$ the expression is slightly
more complicated \cite{Drouffe:1991}
\begin{equation}
  \label{eq:ising_torus_exact}
  \langle\sigma(0,0)\sigma(x,t)\rangle = \frac{\displaystyle \sum_{\nu=1}^4\abs{\vartheta_\nu(\tfrac{z}{2},q)}\abs{\frac{2\pi\eta(i\tau)^3}{\vartheta_1(z,q)}}^{\tfrac{1}{4}}}{\displaystyle \sum_{\nu=2}^4\abs{\vartheta_\nu(0,q)}},\quad z = \frac{\pi}{L}\left(x+ it\right),\quad q = e^{-\pi\tau}
\end{equation}
but the scaling dimension is still present in the exponent $\tfrac{1}{4}=2\Delta$.
$\vartheta_\nu(z,q)$ are the Jacobi theta functions which are quasi-doubly periodic in $z$ and
$\eta(z)$ is the Dedekind eta function. The nome $q$ will often be referred to as a geometry factor
since it depends on the aspect ratio $\tau$. It is worth noting that although the size $L$ explicitly breaks
scale invariance, the correlation function is completely independent of $L$ when expressed in the
dimensionless variables $(\tfrac{x}{L},\tfrac{t}{L})$ (see Fig.~\ref{fig:ising_meff}). Actually, by taking
a small $z$ expansion of the two-point function on the torus, eq.~\eqref{eq:ising_torus_exact}, one obtains the
$\R^2$-result, eq.~\eqref{eq:ising_r2}. However, since a generic model can have distinct fixed
points in the ultraviolet and the infrared it will be useful to study the infrared properties of
eq.~\eqref{eq:ising_torus_exact}.

While eq.~\eqref{eq:ising_torus_exact} is the exact solution it does not allow for an analytic
summation over $x$, required to extract the zero spatial momentum temporal correlator.
However, this is possible if we expand in powers of the geometry factor $q$ (which is small even for
a unit aspect ratio). The resulting correlator is given by
\begin{equation}\label{eq:ising_torus}
  C(t_L,q,L) \propto \cosh\left(\tfrac{\pi}{4L}t_L\right) + \sum_{n=1}^\infty\left\{c_nq^{2n}\cosh\left(m_nt_L\right)+\hat{c}_nq^{n+\tfrac{1}{4}}\cosh\left(\hat{m}_nt_L\right)\right\},
\end{equation}
where $t_L = t-\tfrac{\tau L}{2}$, $m_n=\tfrac{4\pi}{L}(n+\tfrac{1}{16})$, $\hat{m}_n = \tfrac{2\pi}{L}(n-\tfrac{1}{8})$ and, up to $\mathcal{O}(q^2)$,
\begin{align}
  c_n =& \left(\frac{\Gamma(n+\tfrac{1}{8})}{n!\Gamma(\tfrac{1}{8})}\right)^2 +
  \displaystyle \sum_{m=0}^{\floor{\tfrac{2n}{L}}}\frac{2\Gamma(\tfrac{7}{8})^2}{\Gamma(\tfrac{7}{8}-n-\tfrac{mL}{2})\Gamma(\tfrac{7}{8}-n+\tfrac{mL}{2})\Gamma(1+n-\tfrac{mL}{2})\Gamma(1+n+\tfrac{mL}{2})}\\
  \hat{c}_n =& \displaystyle \sum_{m=0}^n\,\sum_{r=\max(0,2m-n)}^m\,\sum_{p=0}^{\mathclap{n+r-2m}}(-1)^r
  \binom{-\tfrac{1}{8}}{m}\binom{\tfrac{1}{2}}{n+r+p-2m}\binom{n+r+p-2m}{2p}\binom{m}{r}\\\nonumber
  &\times\sum_{k=0}^r\binom{r}{k}\left(\binom{2p}{p+r-2k} + 2\sum_{q=1}^{\floor{\tfrac{r+p}{L}}}\binom{2p}{p+r-2k+qL}\right).
\end{align}
We now see clearly that $\tfrac{4\pi}{L}(\tfrac{2\pi}{L})$ plays the role of a mass spacing in $m_n(\hat{m}_n)$and that we obtain
a continuous distribution of masses in the $L\to\infty$ limit. This is even clearer if we
consider the $\tau=\infty$, i.e. cylindrical geometry, and $L\gg1$ limit of the correlator:
\begin{equation}
  C(t,0,L) \propto \sum_{n=0}^\infty \left(\frac{\Gamma(n+\tfrac{1}{8})}{n!\Gamma(\tfrac{1}{8})}\right)^2e^{-\tfrac{4\pi}{L}(n+\tfrac{1}{16})t},\quad \frac{\Gamma(n+\tfrac{1}{8})}{n!} = n^{-\tfrac{7}{8}}\left(1+\mathcal{O}(1/n)\right).
\end{equation}
Going back to Fourier space with the $L\to\infty$ limit in mind one gets
\begin{equation}
  \label{eq:cofw}
  C(\omega,0,L\to\infty)\propto\displaystyle\frac{1}{L}\sum_{n=0}^\infty\left(\frac{L^2}{n^2}\right)^{\tfrac{7}{8}}\frac{8\pi\tfrac{n}{L}}{\left(4\pi\tfrac{n}{L}\right)^2+\omega^2} \approx \int_0^\infty\idb[M]2M\frac{(M^2)^{-\tfrac{7}{8}}}{M^2+\omega^2} = \int_0^\infty\idb[M^2]\frac{(M^2)^{\tfrac{1}{8}-1}}{M^2+\omega^2},
\end{equation}
which is exactly the relation eq.~\eqref{eq:unparticle} for a conformal propagator. This shows how
to obtain a massless correlator with an anomalous scaling dimension as the limit of a model with
a scale. Also, this result casts doubt on the applicability to a cylindrical system of the description~%
\eqref{eq:pole_shift} where the pole of the anomalous propagator is shifted. 

At finite aspect ratio the second infinite sum in eq.~\eqref{eq:ising_torus} contributes more excited
states with half the mass spacing. One finds that unless $\tau$ is larger than around four, these new states
actually lead to a false mass plateau above $\pi/(4L)$ but the correlators still match after a rescaling of
all dimensions which is consistent with the underlying conformal invariance. The effect of a small aspect ratio
on the effective mass can be seen in Fig.~\ref{fig:ising_meff} (\emph{right}).

Away from criticality, the solution of the $2d$ Ising model on a torus is
not available. Therefore, we move to another model where a mass deformation can easily be introduced.
\section{The Sommerfield model}\label{sec:sommerfield}\noindent
The Sommerfield model \cite{Sommerfield:1964}
\begin{equation}
  \mathcal{L} = \bar{\psi}(i\slashed\partial-e\slashed A)\psi - \frac{1}{4}F^{\mu\nu}F_{\mu\nu} + \frac{m_0^2}{2}A^\mu A_\mu
\end{equation}
is the Schwinger model with a mass term for the vector boson. The mass term breaks gauge invariance
but makes it possible to construct a conserved axial current which protects the chiral symmetry.
Since there is no chiral condensate, $\expv{\bar{\psi}\psi}$, the theory remains scale invariant
for any value of the bi-fermion field anomalous scaling dimension between zero and minus one.
This is in contrast to what happens in the one-flavor Schwinger model which stays scale-invariant
due to the bi-fermion field having an anomalous scaling dimension of minus one which yields a dimensionless
condensate. In the Sommerfield model it turns out that the anomalous dimension can be tuned by changing
the dimensionless ratio $\tfrac{\pi m_0^2}{e^2}$.

The model is solved \cite{Sommerfield:1964,Georgi:2010} by introducing a Hodge decomposition
of the gauge field $A_\mu = \partial_\mu \mathcal{V} + \epsilon_{\mu\nu}\partial^\nu\mathcal{A}$
and a rotation of the fermion field $\Psi = e^{ie(\mathcal{V}+\mathcal{A}\gamma^5)}\psi$. In these
new variables the Lagrangian is that of a free fermion and two scalar bosons:
\begin{equation}
  \label{eq:lagrangian_free}
  \mathcal{L} = i\bar{\Psi}\slashed\partial\Psi + \frac{m_0^2}{2}\partial_\mu\mathcal{V}\partial^\mu\mathcal{V}
  +\frac{1}{2}\mathcal{A}(\partial_\mu\partial^\mu)^2\mathcal{A}-\frac{m^2}{2}\partial_\mu\mathcal{A}\partial^\mu\mathcal{A}.
\end{equation}
The shift in the mass $m^2 = m_0^2+\tfrac{e^2}{\pi}$ comes from the change in the measure due
to the $e^{ie\mathcal{A}\gamma_5}$ part of the fermion transformation, just as in the Schwinger
model.

Since the ``new'' fermion $\Psi$ is free it is easy to calculate its $n$-point functions which can then
be dressed by the bosonic contribution to obtain the $n$-point functions of the original
fermion fields $\psi$. We demonstrate the procedure by calculating the two-point function 
\begin{equation}
  \bracket{0}{{\rm T}\psi_\alpha(x)\psi_\beta^*(0)}{0} = \bracket{0}{{\rm T}e^{-ie(\mathcal{V}(x)+\mathcal{A}(x)\gamma^5)}e^{ie(\mathcal{V}(0)+\mathcal{A}(0)\gamma^5)}}{0}\bracket{0}{{\rm T}\Psi_\alpha(x)\Psi_\beta^*(0)}{0}.
\end{equation}
$\Psi_\alpha,\,\alpha\in\{1,2\}$ are the different chiralities of the fermion spinor and we
have $\gamma_5\Psi_1=\Psi_1,\; \gamma_5\Psi_2 = - \Psi_2$. Using the free bosonic propagators which can be read
off from eq.~\eqref{eq:lagrangian_free},
\begin{equation}
  \int\!\!\rd^2x\,e^{ipx}\bracket{0}{{\rm T}\mathcal{V}(x)\mathcal{V}(0)}{0} = \frac{1}{m_0^2p^2},\;
  \int\!\!\rd^2x\,e^{ipx}\bracket{0}{{\rm T}\mathcal{A}(x)\mathcal{A}(0)}{0} =\frac{1}{m^2}\left(\frac{1}{p^2+m^2}-\frac{1}{p^2}\right),
\end{equation}
we obtain
\begin{equation}
  \bracket{0}{{\rm T}\psi_\alpha(x)\psi_\beta^*(0)}{0} = C_0(x)C(x)^{2\delta_{\alpha\beta}-1}S_0^{\alpha\beta},\quad
  S_0^{\alpha\beta}\equiv\bracket{0}{{\rm T}\Psi_\alpha(x)\Psi_\beta^*(0)}{0},
\end{equation}
where
\begin{align}
  C(x) &= \exp\left[\frac{e^2}{m^2}\left(\left(D(x,m)-D(0,m)\right)-\left(D(x,0)-D(0,0)\right)\right)\right],\\
  C_0(x) &=\exp\left[\frac{e^2}{m_0^2}\left(D(x,0)-D(0,0)\right)\right], \quad
  D(x,m) = \int\frac{\rd^2p}{(2\pi)^2}\frac{e^{-ipx}}{p^2+m^2}.
\end{align}
The pseudoscalar $\pi_0$ can be decomposed into the difference of two fermions bilinears,
\begin{equation}
  \label{eq:ps_decomp}
   \pi_0 \equiv \bar{\psi}\gamma_5\psi = \psi_2^*\psi_1-\psi_1^*\psi_2 \equiv \mathcal{O}-\mathcal{O}^*,\quad
  \expv{\pi_0\pi_0^*} = 2\expv{\mathcal{O}\mathcal{O}^*} - 2\Re\left[\expv{\mathcal{O}\mathcal{O}}\right].
\end{equation}
The ``unparticle'' operator $\mathcal{O}$ is composite and related to the product of two fermion operators
\begin{equation}
  \label{eq:unpart}
  \psi_2^*(x_2)\psi_1(x_1) = c(x_1-x_2)\mathcal{O}(x) + \cdots
\end{equation}
through an operator product expansion. In correlators $\expv{\mathcal{O}(t)\mathcal{O}(0)}$ where $t\gg \abs{x_1-x_2}$ the coefficient $c(x_1-x_2)$ can safely be dropped and $\mathcal{O}(t)$ can be considered a local operator. The resulting two-point functions
can be obtained from the corresponding fermionic four-point functions and are found to be
\begin{equation}
  \bracket{0}{{\rm T}\mathcal{O}(x)\mathcal{O}(0)^*}{0} = C(x)^4\abs{S^{12}_0(x)}^2,\quad
  \bracket{0}{{\rm T}\mathcal{O}(x)\mathcal{O}(0)}{0} = C(x)^{-4}S^{12}_0(x)^2.
\end{equation}
The asymptotic behavior of the bosonic correction is
\begin{equation}
  C(x) \propto \begin{cases} 1, & xm\ll 1\\
    (x^2)^{-\gamma_\mathcal{O}/4}, & xm\gg 1 \end{cases},\quad \gamma_\mathcal{O} = -\frac{e^2}{\pi m^2} =
  -\frac{1}{1+\tfrac{m_0^2\pi}{e^2}},
\end{equation}
which makes it clear that $\gamma_\mathcal{O}$ is the anomalous dimension of $\mathcal{O}$ in the infrared.
It can be varied between 0 for the free case, $e=0$, and $-1$ for the Schwinger case, $m_0=0$. Since
$\bracket{0}{{\rm T}\mathcal{O}(x)\mathcal{O}(0)}{0}$ decays faster than
$\bracket{0}{{\rm T}\mathcal{O}(x)\mathcal{O}(0)^*}{0}$ by a factor of $(x^2)^{2\gamma_\mathcal{O}}$
(even exponentially on a finite lattice)
the pseudoscalar correlator will be dominated by the latter and we will focus on it from now on.
%On a finite lattice the supression will even be exponential due to the masses induced by the finite $L$.

\subsection{On the lattice}\label{sec:latt}\noindent
To study the Sommerfield model on a lattice of size $L_s\times L_t = (a N_s)\times(a N_t)$ we
substitute the continuous momentum $p$ with its lattice counterpart
\begin{equation}
  \label{eq:lattmom}
  a\tilde{p}_{s,t} = \sin\frac{2\pi}{N_{s,t}}(n+q_{s,t}),\quad a\hat{p}_{s,t} = 2\sin\frac{\pi n}{N_{s,t}},
  \quad n \in \{0,1,2,\ldots N-1\}, \quad q_{s,t} \in \{0,\tfrac{1}{2}\}
\end{equation}
for fermions and bosons respectively, and all integrals with finite sums. The value of $q_{s,t}$
determines the boundary conditions of the fermions. In the spatial direction they can be either
periodic (pbc, $q_s=0$) or anti-periodic (apbc, $q_s=\tfrac{1}{2}$) whereas in the temporal
direction only apbc, i.e. $q_t = \tfrac{1}{2}$, is allowed. The main effect of the spatial boundary
conditions concerns the lower bound on the spatial momentum, $\abs{\tilde{p}_s} \geq 2\pi q_s/L_s$
which, in the case of apbc, leads to a nonzero mass gap as that seen in the Ising model.

\subsection{Results}\label{sec:res}
\begin{figure}
  \centering
    \mbox{\includegraphics[width = 0.45\textwidth]{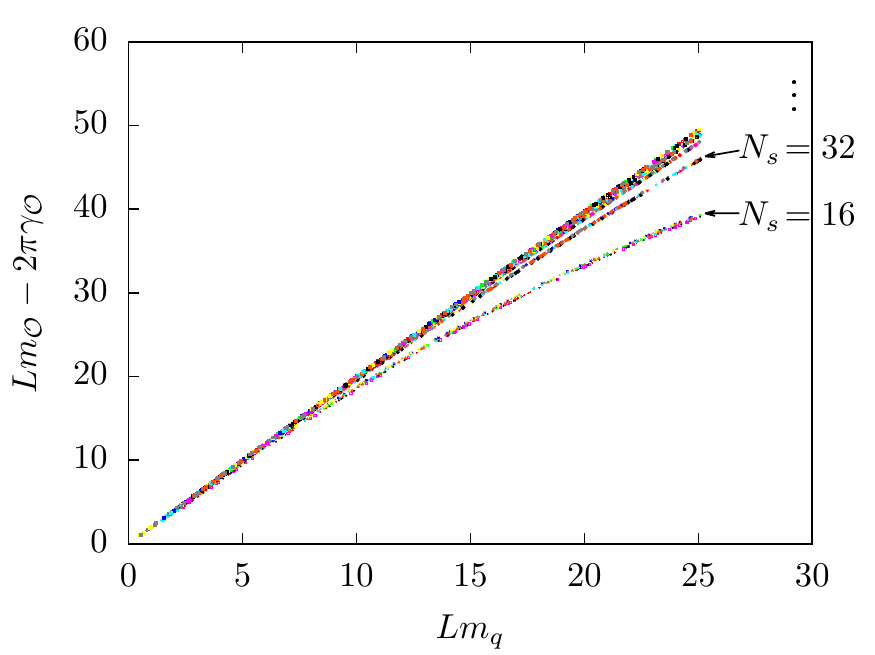}%
    \includegraphics[width = 0.45\textwidth]{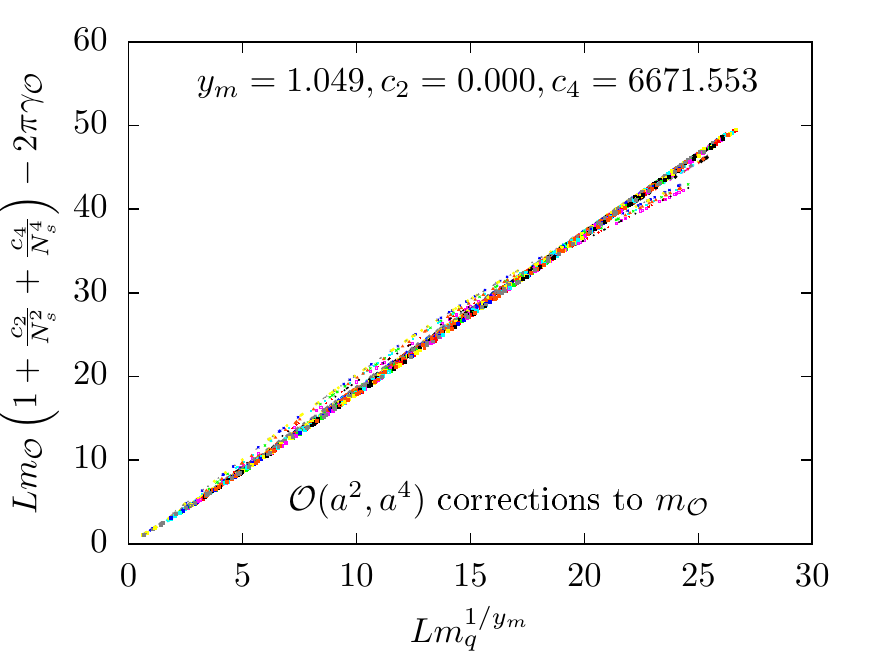}}\\
    \mbox{\includegraphics[width = 0.45\textwidth]{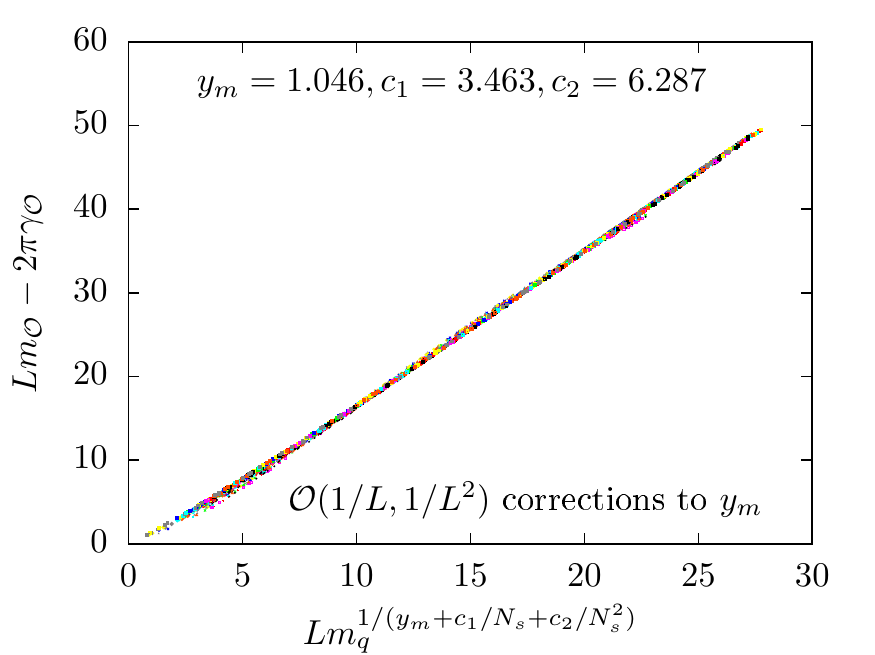}%
    \includegraphics[width = 0.45\textwidth]{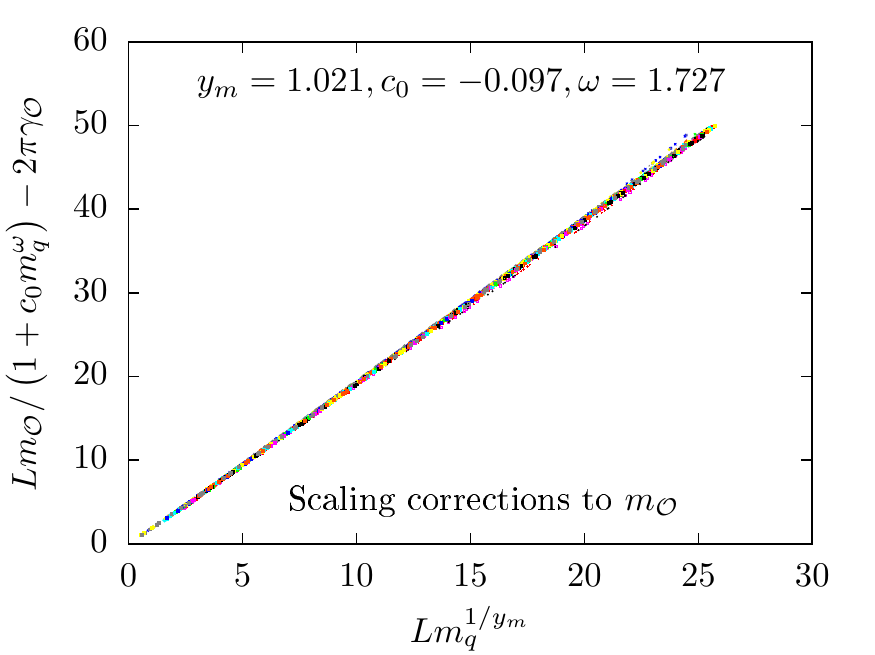}}
    \caption{(Shifted) ``unparticle'' mass, $m_\mathcal{O}$, as a function of the bare mass deformation, $m_q$, and three different strategies to improve the data collapse. Note that the exact value of
the anomalous mass dimension is $y_m=1$. The results are obtained with periodic boundary conditions.}
    \label{fig:sommer_scaling}
  %\end{center}
\end{figure}

\noindent
If a fermion mass term is added to the Sommerfield Lagrangian chiral symmetry is explicitly broken
but the method of solution remains the same, one just has to replace the massless fermion
propagators by their massive counterparts. With this observation we can measure the ``unparticle''
mass as a function of the mass deformation. First, the mass is shifted from zero to $2\pi\gamma_\mathcal{O}$ (cf. Ising model with
$\Delta=1+\gamma$). Furthermore, we can extract the anomalous mass dimension $\gamma_m$ through
the relation $Lm_\mathcal{O}  \propto Lm_q^{1/y_m}$ where $y_m=1+\gamma_m$ is the
total mass dimension. The results with $y_m=1$ fixed at its exact value\footnote{This value can most easily be
derived by considering the point-to-point correlator, esp. the condensate, and noticing that the $m_q$ dependence
is that of a free fermion model.}
are displayed in the upper
left panel of Fig.~\ref{fig:sommer_scaling}. Clearly, the data are not well described by the
naive scaling ansatz. Our first attempt at improving the data collapse is to consider discretization
errors of $m_\mathcal{O}$ (\emph{upper right panel}). Another approach is to consider an
$N_s$-dependent anomalous dimension which runs towards its fixed point value as $N_s\to\infty$
(\emph{lower left panel}). Yet another approach is to consider corrections to scaling. In this
case the naive scaling relation is changed to include also the first irrelevant operator and
the result is displayed in the lower right panel of Fig.~\ref{fig:sommer_scaling}. It is evident
that all three methods greatly improve the data collapse and the resulting anomalous dimensions
are close to the exact result. \pagebreak It is not obvious which method is the correct one but the issue
could possibly be settled by calculating the next irrelevant operator and comparing its exponent
with the fitted $\omega$.

\section{Conclusions}\label{sec:conclusions}\noindent
Although a finite lattice explicitly breaks conformal invariance some of the scale invariance is
still encoded in the lattice correlator which is left invariant when all dimensions are
rescaled. This is true also when the finite lattice extent generates an exponentially decaying, massive correlator. We have
also demonstrated how the full conformal correlator is obtained in the infinite volume limit from
a continuous distribution of massive states.

In the presence of a mass deformation we find that the extraction of the mass anomalous dimension
can be subject to some ambiguity when it comes to subleading effects.

\end{document}